%% file: Ad_Pump_8.tex
\documentclass[aps,prl,twocolumn,superscriptaddress]{revtex4-1}

\usepackage{graphicx}  
\usepackage{amsmath} 

\usepackage{graphicx}
\usepackage{amsmath,amssymb,amsfonts}
\usepackage{textcomp}
\usepackage{hyperref}
\usepackage{gensymb}
\usepackage{verbatim}
\usepackage{braket}

\hypersetup{breaklinks=true,colorlinks=true,urlcolor=black}
\usepackage{color}
\usepackage{soul}
\usepackage{gensymb}

\DeclareGraphicsExtensions{.jpg,.pdf,.png,.eps}

\begin{document}


\title{Geometric pumping and dephasing at topological phase transition}



\author{B.Q.~Song}
\affiliation{Ames Laboratory, Iowa State University, Ames, Iowa 50011, USA}
\affiliation{Department of Physics and Astronomy, Iowa State University, Ames, Iowa 50011, USA}
\author{J.D.H.~Smith}
\affiliation{Ames Laboratory, Iowa State University, Ames, Iowa 50011, USA}
\affiliation{Department of Mathematics, Iowa State University, Ames, Iowa 50011, USA}
\author{L.~Luo}
\author{J.~Wang}
\affiliation{Ames Laboratory, Iowa State University, Ames, Iowa 50011, USA}
\affiliation{Department of Physics and Astronomy, Iowa State University, Ames, Iowa 50011, USA}


\date{\today}

\begin{abstract}
A measure-preserving formalism (MPF) is constructed and applied to spin/band models, which yield observations about pumping. It occurs at topological phase transition (TPT), i.e., a $Z_2$-flip, suggesting that $Z_2$ can imply bulk effects. The model's asymptotic behavior is analytically solved via MPF. The pumping probability is geometric, fractional, and has a ceiling of $\frac{1}{2}$. Intriguingly, theorems are proved about occurrence conditions, which are linked to the system's dimension and the distinction between rational and irrational numbers. Experimental detection is discussed.
\end{abstract}



\maketitle



As witnessed in the past decades, geometry has intertwined with physics \cite{Moody, Niu} primarily via a framework outlined by Berry \cite{Berry} and Simon\cite{Simon}, which has two crucial ingredients: one is adiabaticity, vital for the analogy to ``parallel transport”; the other is Berry curvature which makes anholonomy occur. The two points are believed necessary, serving the foundation of recognizing the slow/fast variables, deducing topological invariants, etc. as seen in quantizatization of adiabatic pumping \cite{Thouless, Fu} or transports \cite{TKNN}, theory of electric polarization \cite{Vanderbilt}, topological gaped \cite{TISC, Symm} or nodal states \cite{Weyl}.



Our first result is to incorporate geometry without adiabaticity and Berry curvatures – the two ingredients are proved dispensable. Our scheme relies on making quantum evolution mimic a classical trajectory via a measure-preserving formalism (MPF) \cite{MPF, Statistics}, which originates from efforts in unifying formalisms for quantum and classical mechanics \cite{Wigner, Flow, Gro46, Moy49}. In Berry’s scheme (and its development \cite{Samuel, AA}), local curvature is the key; here, the ``trajectory” in phase space is at heart, entailing different geometric intuitions (e.g., ergodicity \cite{Ebook}) and math apparatus. Moreover, MPF helps deduce exact solving of a spin model to establish a rigorous concept from which one can set out to examine general situations.



The second result is about an observable, an inter-band pumping, namely \textit{geometric pumping}. Its rate purely depends on an angle parameter, which, in certain instances, is linked to $Z_2$ index \cite{FuInv, FieldTI}. Since the pumping is for bulk, it challenges the wisdom that $Z_2$ only implies surface/interface, known as bulk-edge correspondence \cite{Bernevig}. It relies on the flipping of $Z_2$ (not a specific state of $Z_2$) thus is a genuine consequence of topological phase transitions (TPT). Its detection perfectly matches the pump-probe laser technique \cite{Chirag, Liang}: ``pump" is to excite phonons that can distort the bands and may induce electronic TPT, and ``probe" is to detect charge pumping. Suitable systems include narrow-gap topological insulators or semi-metals, which are liable to undergo TPT. A candidate recently studied is ZrTe$_5$, whose topological states can be altered by A$_{1g}$ \cite{Chirag}, B$_{1u}$ modes \cite{Liang}.

The work spans several subjects: topological band theory, MPF, ergodic theory, measure theory, number theory. It is fascinating that such seemingly irrelevant topics appear together without a planner. We protrude their relations in the main text and leave backgrounds, derivation, and proof in Appendix and references.

\begin{figure}
\includegraphics[scale=0.3]{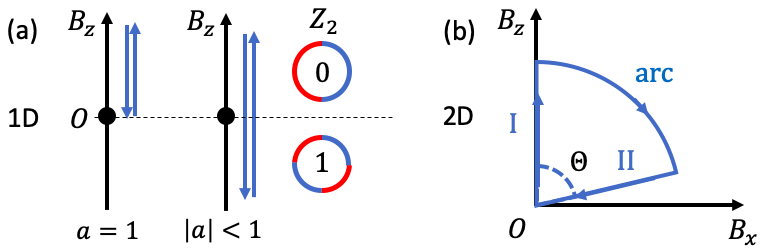}
\caption{\label{fig:epsart}(color online): (a) 1D. $Z_2:0{\rightarrow0}$, no pumping; $0{\rightarrow1}$, pumping. (b) 2D. \textbf{B}$(t)$ follows a sector loop with angle $\Theta$. Dimensions refer to the spatial ones. For example, if $\textbf{B}(t)$ is restricted along the $z$-axis, it is 1D; if allowed for rotation in a plane, e.g., the $x$-$z$ plane, it is 2D.\label{f1}}
\end{figure}

We set out from a spin model to introduce a generic MPF. 
Consider a spin-$\frac{1}{2}$ in 1D cyclic magnetic field $\textbf{B}(t)=B_{0}(a+\text{cos}({\omega}t)){\cdot}{\hat{z}}$, two levels with a gap $\Delta(t)=2{\mu}_{B}B_{z}(t)$ which will momentarily close if $|a|{\leq}1$ (Fig.~\ref{f1}(a)). Unlike previous models constantly with gaps \cite{Berry, Thouless, Fu, Zee}, we allow gap-closing (thus non-adiabatic) for investigating TPT. $Z_2$ index can be defined as $\frac{1}{\pi}\text{cos}^{-1}(\text{sgn}({B}_z{\cdot}\sigma_z))$, valued 0 or 1 depending the relative orientations of $\textbf{B}$ and spin. Flipping $Z_2$ requires $\textbf{B}$ passing zero, i.e., gap closing. 

The 1D setting is to introduce the concept of “pumping,” which refers to the spin being excited to the higher eigenstate ${|n_1{\rangle}}$ (${|n_0{\rangle}}$ is ground state) under the influence of $\textbf{B}(t)$. When generalized to band models, this corresponds to inter-band pumping, different from pumping across real space onto edges \cite{Thouless, Fu, Moore}. For 1D, the exact solution is available, say starting with spin up: 
\begin{equation}
\begin{split}
|{\varphi}(t){\rangle} = e^{-i{\int}^{t}E_{\uparrow}(\tau)d{\tau}}|{\uparrow}{\rangle},
\end{split}
\label{eq1}
\end{equation} 
where $|{\uparrow}{\rangle}$ stands for the state of spin up and $E_{\uparrow}(\tau)={\langle}{\uparrow}|H(\tau)|{\uparrow}{\rangle}=-{\mu}_{B}B_{z}(t)$. So $|{\varphi}(t){\rangle}$ sticks to $|\uparrow{\rangle}$. But $|\uparrow{\rangle}$ might switch between $|n_0{\rangle}$ and $|n_1{\rangle}$, when gap is closed (Fig.~\ref{f1}(a)). If level-inversion happens ($Z_2$ is flipped), pumping occurs; if $Z_2$ does not change, e.g., $a{\geq}1$, no pumping takes place. 
 
Despite its simplicity, 1D clearly shows geometric characteristics: (1) pumping only depends on the flipping of $Z_2$, insensitive to eigenvalues unless the gap is closed; (2) Eq.~\ref{eq1} holds for arbitrary $\omega$ and lacks energy scales (e.g., ${\Delta}/{\omega}$). Eq.~\ref{eq1} indicates, after two cycles, the spin will be de-pumped back to the original eigenstate up to a phase. Such returnable behavior is liable to be generic for 1D, as it is also observed in 1D $Z_2$ spin pump \cite{Fu}. In that case, a spin chain under cyclic changing potentials is considered, and after one cycle, the spin is pumped from one edge to the other, but in the next cycle, the pumped spin will return (i.e., net pumping in two cycles is zero). Thus, external dephasing (e.g., coupled with leads) is believed necessary to crush coherence and make pumping happen \cite{Fu, Zhou}. However, when the dimension is ${\geq}2$, as we find below, ``internal" dephasing occurs.

Consider $\textbf{B}(t)$ in $x$-$z$ plane (Fig.~\ref{f1}(b)). The arc section with large $|\textbf{B}|$ ensures ideal adiabatic evolution. Two straight sections with exact solutions are to handle the gap closing. Average pumping rate over $n$ cycles is
\begin{equation}
\begin{split}
p_n=\frac{1}{n}{\sum}_{j=1}^{n}|{\langle}n_1|\mathcal{U}^j|{\varphi}(t=0)\rangle|^2.
\end{split}
\label{eq2}
\end{equation}
Evolution operator $\mathcal{U}({\Theta}, {\Omega}, {\Phi})$ is found (Appx. A), where $\Theta, \Omega$ are polar and azimuthal angles of $\textbf{B}(t)$; $\Phi$ contains dynamic phases from $B_0$ and $\omega$.
\begin{equation}
\begin{split}
\mathcal{U}=\begin{pmatrix} \text{cos}({\Theta}/{2})e^{-i{\Phi}} & -\text{sin}({\Theta}/{2})e^{-i({\Omega}-{\Phi})} \\ \text{sin}({\Theta}/{2})e^{i({\Omega}-{\Phi})} & \text{cos}({\Theta}/{2})e^{i{\Phi}} \end{pmatrix}.
\end{split}
\label{eq3}
\end{equation}
Here, $\mathcal{U}$ will not determine a state midst a loop but will give the resultant state after $n$ loops by $\mathcal{U}^n$. $\mathcal{U}$ being SU(2) rather than U(2) is due to time-reversal $\mathcal{T}H(t)\mathcal{T}^{-1}=-H(t)$ (Appx. A). $\mathcal{U}$ is \textit{independent} of gauge, as required by an evolution operator. Pumping takes place at the pathway corner (gap-closing point), but phases accumulated elsewhere influence. Previous results about pumping \cite{Thouless, Fu, Moore, Zhou} rely on the adiabatic limit $\omega{\ll}\Delta$. Our result is interesting on account of being valid for $\omega{\gg}{\Delta}_{\text{min}}=0$, and error in polynomials $\mathcal{U}^n$ only coming from arc sections, which can be exponentially suppressed by raising $B_0$.

We seek $p_{n{\to}{\infty}}$, limiting behaviors after a number of cycles. Later we will show $p_{n{\to}{\infty}}$ converges almost everywhere (a.e.) in $\Theta$-$\Phi$ space (also $\Theta$-$\Phi$-$\Omega$ for 3D). It is straightforward to show $p_n(\Theta)=p_n(-\Theta)$ and $p_n(\Phi)=p_n(\Phi+\pi)$, so we adopt $\Phi{\in}[-\pi/2, \pi/2]$ and $\Theta{\in}[0, \pi]$. For 1D, $\Phi$ is the variable and $\Theta=\pi$ or 0, and $p_n$ is integer for arbitrary $\Phi$. For 2D ($0<\Theta<\pi$), ``fraction" pumping emerges, e.g., $\frac{1}{2}$ ``shot" of spin is pumped when $x$- is projected to $z$-axis. Projections occur in every cycle, subject to computation of $p_{n}$ with Eq.~\ref{eq2}, which becomes difficult as $n$ getting large. Fortunately, an ingenious method of MPF finds $p_{\infty}$ converge to analytic expressions.

MPF is built for both classical and quantum, but differently. For classical,  preservation of measure  $dp~{\wedge}~dq$ is directly deduced from Hamilton's equations \cite{Statistics}, thus MPF is established on rigor. For quantum, however, 
measure preservation demands all higher-order ($>$2) derivatives of potentials must be vanishing \cite{Wigner, Flow}. Thus, the validity of MPF relies on stringent conditions. This motivates us to construct MPF on an equal footing: Just as Liouville's Theorem follows from Hamilton's equations, can measure-preservation result directly from Schr\"odinger equation (SE)? One feature of SE (perhaps unique) that ensures measure preservation is unitarity of the evolution. A unitary operator is an endomorphism of Hilbert space, and continuous endomorphism of a compact group will preserve its Haar measure \cite{Ebook}. Accordingly, two modifications are made. First, we leave the ${\lbrace}p, q{\rbrace}$ space and turn to the group's parameter space. Second, Haar measure (Appx. B) defined for the symmetry group on Hilbert space $\mathcal{H}$ is to replace the measure $dp~{\wedge}~dq$. 

In Hamilton mechanics, evolution operator $T_t$ updates particles' momenta and positions \cite{Statistics, Ebook}; formally, it is a set ${\lbrace}T_t|t\in\mathbb{R}{\rbrace}$ that transform the space ${\lbrace}p, q{\rbrace}$, preserving the symplectic measure $dp~{\wedge}~dq$ (Liouville theorem), and satisfying semigroup law $T_{t+s}=T_t{\circ}T_s$. Taking discrete time steps $s$, the semigroup law yields $T_t=T^n_s$ for $t=ns$. In comparison, quantum $\mathcal{U}$ represents the evolution during $\textbf{B}(t)$ completing a loop, serving as a ``time step". $\mathcal{U}$ is a matrix of SU(2) group, satisfying $\mathcal{U}^{l+n}=\mathcal{U}^{l}{\circ}~\mathcal{U}^{n}$. It will rotate $\mathcal{H}$ and preserve SU(2)'s Haar measure $m(\Theta, \Omega, \Phi)=2\text{sin}(2\Theta){\cdot}d{\Phi}{\cdot}d{\Theta}{\cdot}d{\Omega}$ (also its subgroups'). We observe the correspondence:
\begin{equation}
\begin{split}
T_s^n(p, q) \sim \mathcal{U}^n(\Theta, \Omega, \Phi);~~dp~{\wedge}~dq\sim m(\Theta, \Omega, \Phi).
\end{split}
\label{eq4}
\end{equation}
The step number $n$ is ``time". The phase-space coordinates $(p, q)$ correspond to 3-dimensional state coordinates $(\Theta, \Omega, \Phi)$, referring to the spin orientation and phases. The symplectic measure $dp~{\wedge}~dq$ corresponds to the Haar measure $m$ (Appx. B). The correspondence for the time averages for an observable $f$ is given by
\begin{equation}
\begin{split}
\bar{f}=\frac{1}{t}\operatorname*{\sum}_{i=0}^{n}f(T_s^i(p, q))~{\sim}~\bar{f}=\frac{1}{n}\operatorname*{\sum}_{i=0}^{n}f_np_n,
\end{split}
\label{eq5}
\end{equation}
where $f$ is certain observable to be averaged. By the substitution of Eq.~\ref{eq4} and \ref{eq5}, we define a MPF. 

The valid conditions for MPFs are summarized in Table~\ref{tab1}. There is an elegant correspondence between the Liouville theorem and the present scheme. Liouville requires ``Hermitian" in a classical sense, i.e, the system's dynamics obey Hamilton equations $\dot{q}={\partial}_pH, \dot{p}=-{\partial}_qH$ \cite{Statistics}. However, if dissipative terms exist (e.g., $\dot{p}=-{\partial}_qH-p$), MPF is invalid. For this work, it requires ``Hermitian" in a quantum sense, i.e., Hamiltonian is a Hermitian operator $H^{\dagger}(t)=H(t)$, which is a fundamental feature of quantum mechanics that ensures evolution $\mathcal{U}$ being unitary, deducing robust MPF. In both MPFs, $H$ can be time-dependent, thus energy is allowed to flow in or out. On the other hand, Wigner's scheme requires perfect harmonic potentials \cite{Wigner}, which are often not satisfied even at an approximation level.

\begin{table}
\caption{\label{tab:table1} Comparion of different MPFs: Liouvile theorem \cite{Statistics}, Wigner flow \cite{Wigner} and this work,  in terms of applied scopes (classical or quantum), measure functions, and valid conditions.\label{tab1}}
\begin{ruledtabular}
\begin{tabular}{c c c c}
MPF & Scope & Measure & Condition \\
\hline
Liouville & C & $dp~{\wedge}~dq$ & $\dot{q}={\partial}_pH, \dot{p}=-{\partial}_qH$ (Robust) \\
Wigner & Q & $dp~{\wedge}~dq$ & $\partial_q^nH=0$ for $n>2$ (Fragile) \\
This work & Q & Haar &  $H^{\dagger}=H$ (Robust) \\
\end{tabular}
\end{ruledtabular}
\end{table}

To proceed, we need a crucial concept: ergodicity (intuitive interpretation seen in Appx. B). A system being ergodic means it can reach every region in the space (either ${\lbrace}p, q{\rbrace}$ space or others) after sufficient time. Formally, it is a property of the evolution operator.

\textbf{Definition 1.} Let ($X$,$\mathfrak{B}$,$m$) be a probability space. A measure-preserving transformation $T$ of ($X$,$\mathfrak{B}$,$m$) is called ergodic if the only members of $B$ of $\mathfrak{B}$ with $T^{-1}B=B$ satisfy $m(B)=0$ or $m(B)=1$.

Here, $\mathfrak{B}$ denotes a $\sigma$-algebra of set $X$. The $m$ is a measure function that is equipped to depict ``probability" and has been normalized to 1 (Appx. B). If $T$ is ergodic, we have the following theorems (p. 30, 34 of \cite{Ebook}). Theorem 1.2 is known as Birkhoff Ergodic theorem.

\textbf{Theorem 1.1}. Let $G$ be a compact group and $T(x)=ax$ a rotation of $G$. Then $T$ is ergodic iff ${\lbrace}a^n{\rbrace}^{\infty}_{-\infty}$ is dense in $G$. In particular, if $T$ is ergodic, then $G$ is abelian. 

\textbf{Theorem 1.2}. Let $T$:($X$,$\mathfrak{B}$,$m$)$\rightarrow$($X$,$\mathfrak{B}$,$m$) be measure preserving and $f{\in}L^1(m)$. Then $(1/n){\Sigma}_{i}^{n-1}f(T^i(x))$ converges a.e. to a function $f^*{\in}L^1(m)$. Also $f^*{\circ}T=f^*$ a.e. and if $m(X)<\infty$, then ${\int}f^*dm={\int}fdm$. 

Theorem 1.1 implies that the non-abelian SU(2) cannot be ergodic but can admit an ergodic abelian sub-group, when ${\lbrace}T^n{\rbrace}_{\infty}$ is dense, i.e., $T^n{\neq}\mathbb{I}$ for $n{\in}\mathbb{N}$. Theorem 1.2 implies that if $T$ is ergodic, an auxiliary function $f^*$ can be introduced to evaluate the (Lebesgue) integration of the original $f$. The $f^*$ gives the probabilities of occupying a region in phase space. $f^*{\circ}T=f^*$ renders a property similar to translation invariant, except for the set ${\lbrace}T^n{\rbrace}_{\infty}$ is dense rather than continuous. Accordingly, $f^*=\rho$ a.e. rather than everywhere. Theorem 1.2 premises $m<\infty$. Thus, finite $m$ is indispensable.

\begin{figure}
\includegraphics[scale=0.33]{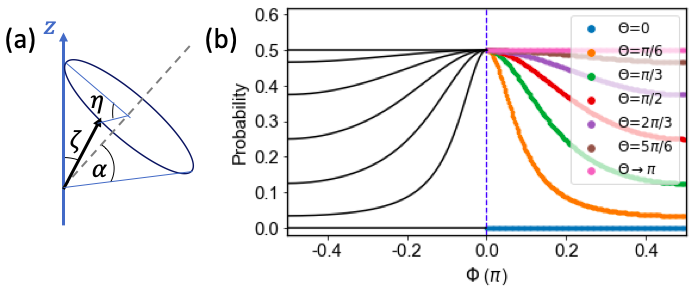}
\caption{\label{fig:epsart}(color online): (a) Geometric pumping casued by ``spin rotation". Note the rotation is not smooth, but jumping by angle ${\Delta}\eta=\delta$ each loop, and $\delta$ is given by Eq.~\ref{eq17}. With ergodicity (unstable points), $\delta/\pi$ is irrational, and no points are coinciding. With $n\to\infty$, it will form a quasi-continuous trajectory composed by dense pionts. (b) Numerical results (colored) against analytic solution (black). \label{f2}}
\end{figure}

The ergodic subgroup is explicitly found (Appx. C) and has a geometric interpretation (Fig.~\ref{f2}(a)): a spin vector rotates in a ``trajectory" around a fixed axis, and pumping is simply the projection to $-z$. Integration with $f^*=\rho$ over the trajectory (Fig.~\ref{f2}(a)) gives the below analytic result, which perfectly matches numerical evaluations with Eq.~\ref{eq2} (Fig.~\ref{f2}(b)).
\begin{equation}
\begin{split}
p_{\infty}=\frac{1}{2}\frac{\text{sin}^2(\frac{\Theta}{2})}{1-\text{cos}^2(\frac{\Theta}{2})\text{cos}^2(\Phi)}.
\end{split}
\label{eq6}
\end{equation}
What is the efficient way to pump spin? Large $\Theta$  pumps more each ``shot", but de-pumping is also more; small $\Theta$ pumps little each time, but accumulation of $n{\to}{\infty}$ is unclear. Eq.~\ref{eq6} gives exact answers to this wondering. In above, $\Theta$ is a pure angle, but $\Phi$ contains dynamic phases. Why Eq.~\ref{eq6} is geometric? Note that $p_{\infty}$ curves get flat as $\Theta{\to}0$ or $\pi$, finally constant with $\Phi$ (Fig.~\ref{f2}(b)). Later we will see $\pi$ and 0 are the physically achievable values for $\Theta$. Thus, pumping is insensitive to dynamic details (e.g., $\textbf{B}(t)$ or $\omega$), unless gap-closing is touched, at which $\Theta$ hops from 0 to $\pi$ or reversely, leaving a purely geometric effect. It is more evident when $p_G=\frac{1}{\pi}\int_{-\pi/2}^{\pi/2}p_{\infty}{\cdot}d\Phi$ is defined. Since $B_0$ is huge, even a small fluctuation will cause a drastic change in phases, making $\Phi$ random statistically. Thus, $p_G$ is the quantity practically linked to observables. If $\Phi$ evenly distributes, analyticity survives
\begin{equation}
p_G=\frac{1}{2}\text{sin}(\Theta/2).
\label{eq7}
\end{equation}
In this case, instead of parallel transport \cite{Moore, Berry}, it is ergodicity that generates a quasi-continuous trajectory (Fig.~\ref{f2}(a)) and lets geometry come in. Eq.~\ref{eq7} rises from ``0/0", i.e., both $\omega$ and $\Delta_{\text{min}}{\to}0$, where $\omega/{\Delta}_{\text{min}}$ is ill-defined. This reminds us of quantum criticality \cite{Sachdev}, a situation of both $\hbar{\omega}{\to}0$ and $k_BT_c{\to}0$, where characteristic energy fails and scale invariance emerges \cite{Sachdev, Son}. Here, geometry emerges. The $\frac{1}{2}$ factor in Eq.~\ref{eq7} is noteworthy. First, it suggests geometric pumping is fractional, different from pumps driven by photons that allow a ``whole" particle to be excited once energy quanta are matched. Second, $p_G$ has a ceiling of $\frac{1}{2}$, a reminiscent of the principle of maximum entropy, because $\frac{1}{2}$ makes $S{\to}S_{\text{max}}=\text{ln}2$ for two levels. Besides, the pumping will be quantized if $\Theta$ is quantized (as shortly seen in the band model). A major discovery here is that the returnable behavior is fragile against dimension perturbation, which is much unnoticed for restriction to perfect 1D \cite{Thouless, Fu, Zhou} or cyclic evolution \cite{AA}. A cyclic spin-rotation considered by Aharonov and Anandan \cite{AA} corresponds to a special case $\Theta=\pi$ here.

\vskip 1mm
\textbf{Definition 2}: The point $({\Theta}, {\Phi}, \Omega)$ is \textit{stable} of order $N$ for a natural number $N$, if $\mathcal{U}^N(\Theta, \Phi, \Omega)$ is diagonal. If the point is not stable for arbitrary $N$, it is \textit{unstable}. 
\vskip 1mm
Note that pumping \textit{only} takes place at unstable points, because the Hamiltonian specified by stable $({\Theta}, {\Phi}, {\Omega})$ allows spin to return to the original state within finite cycles. This is reflected by that $p_{\infty}$ diverges at a stable point. Because, if the sequence includes every $n\in\mathbb{N}$, $p_{\infty}$ is given by Eq.~\ref{eq6}; if $n=mN (m\in{\mathbb{N}})$ and $N$ is the order of stable point, we obtain a subsequence ${\lbrace}p_n{\rbrace}$ to make $p_{\infty}=0$, a distinct result from Eq.~\ref{eq6}. In fact, $p_{\infty}$ can yield many different values by choosing ${\lbrace}p_n|n~\text{mod}(N)=l{\rbrace}$. and $l=0,1, 2...$ Thus, Eq.~\ref{eq6} only converges for unstable points (ergodic). The existence of the limit follows from Theorem II.11 of \cite{Funct}. For stable points of order $N$, spin hops among $N$ states and is non-ergodic. 

We prove two theorems (Appx. D) about the physical conditions for pumping to occur.
\begin{figure}
\includegraphics[scale=0.5]{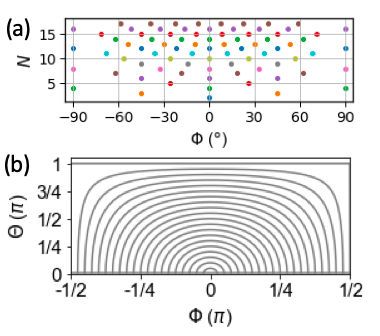}
\caption{\label{fig:epsart}(color online): (a) Stable points up to order $N=17$ for $\Theta=\pi/2$. If $N\to\infty$, stable points form a dense subset over $\Phi$. (b) Global view of stable points in $(\Theta, \Phi)$ phase space. The plotted lines are a portion of the stable curves, which are actually super-dense. (a) can be generated by a cutting line at $\Theta=\pi/2$. The points on boundaries (i.e., $\Phi={\pm}\pi/2$ and $\Theta=0, \pi$) are stable. \label{f3}}
\end{figure}

\textbf{Theorem 2.1}. Let $A$ (resp. $\bar{A}$) denotes the set of stable (resp. unstable) points $(\Theta, \Phi)$.
\vskip 0.5mm
(a) Points with $\Theta\in\{0,\pi\}$ or $\Phi=\pm\pi/2$ are stable.
\vskip 0.5mm
(b) Fix $\Theta$ in the open interval $(0,\pi)$. Then:
\vskip 0.5mm
${~~~~~}$(i) $\Set{\Phi|(\Theta, \Phi)\in A}$ is dense in $[-\pi/2, \pi/2]$.
\vskip 0.5mm
${~~~~}$(ii) $\Set{\Phi|(\Theta, \Phi)\in \bar{A}}$ is dense in $[-\pi/2, \pi/2]$.
\vskip 0.5mm
(c) Fix $\Phi$ in the open interval $(-\pi/2,\pi/2)$. Then:
\vskip 0.5mm
${~~~~~}$(i)$\Set{\Theta|(\Theta, \Phi)\in A}$ is dense in $[0,\pi]$.
\vskip 0.5mm
${~~~~}$(ii)$\Set{\Theta|(\Theta, \Phi)\in \bar{A}}$ is dense in $[0,\pi]$.
\vskip 1mm

\textbf{Theorem 2.2}. Let $(\Phi)$, $(\Theta, \Phi)$ and $(\Theta, \Omega, \Phi)$ denote the points in 1D, 2D and 3D phase space.

\vskip 0.5mm
(a) In 1D ($\Theta=0$, $\pi$), $\bar{A}={\emptyset}$, i.e., $m(\bar{A})=0$, $m(A)=1$.
\vskip 0.5mm
(b) In 2D or 3D, almost every $({\Theta}, {\Phi})$ or $({\Theta}, {\Omega}, {\Phi})\in\bar{A}$, i.e., for Lebesgue measure $m(\bar{A})=1$, $m(A)=0$.
\vskip 1mm
The theorems can be translated into three physical conditions/features for geometric pumping: 
\vskip 1mm
(i) ~Gap closing and TPT. 

(ii) Robustness to energetic/dynamic details, e.g., band gap sizes, driving frequencies $\omega$.

(iii) System's dimension $D>1$. 
\vskip 1mm
Reasoning is as follow. For (i), the vertex of angle $\Theta$ is at the degeneracy point $\textbf{B}=0$ (Fig.~\ref{f1}(b)), thus gap closing is required. Then, from Eq.~\ref{eq7}, $\Theta=\pi$ \cite{Note} gives the maximum $p_G=1/2$, while $\Theta=0$ leads to $p_G=0$. In band models (where $\Theta$ take discrete values $\pi$ or $0$), $\Theta=\pi$ corresponds to band inversion, altering topological states. Thus, TPT is a necessary condition. 

For (ii), the energetic/dynamic information is encoded in $\Phi$ and $\Theta$, i.e., given a $H(t)$, it will project an image in $(\Theta, \Phi)$ space. Pumping occurrence depends on whether the model's regime (it is a finite region, for real systems must have a spread) can touch the distributed areas of $\bar{A}$. Theorem 2.1(b)(c) reveals an interesting pattern: Any finite (compact) regime must contain both $A$ and $\bar{A}$, as they are both dense subsets. This resembles the distribution of rational $\mathbb{Q}$ and irrational numbers $\bar{\mathbb{Q}}$ on the real axis: any finite interval $\mathbb{R}$ must contain $\mathbb{Q}$ and $\bar{\mathbb{Q}}$. Such distribution entails robustness for pumping, because no matter how to adjust the model's parameter (e.g., by changing gap sizes, or the way approach gap-close), no matter what size or shape of the regime the model occupies in the $(\Theta, \Phi)$ space, encountering $\bar{A}$ is unavoidable - all because $\bar{A}$ is so densely embedded. Remarkably, such robustness is endowed by an unprecedented source of math. We already know the protection for a physical effect can be established on symmetry group or topology theories \cite{TISC, Symm, Weyl}. Here the protection is rendered by facts in number theory. In fact, this robustness is independent of symmetry, which is a rare virtue since rigor often breaks down for lacking required symmetries. For example, the Mermin-Wagner theorem (the absence of long-range order in low-dimensional systems at finite temperatures) relies on isotropic presumption and only leads to a tentative argument in crystals, where the needed continuous symmetry is absent. 

Moreover, the distribution of $A$ and $\bar{A}$ can be solved and analytically expressed. In Appx. C, we find $A$ ($\bar{A}$) is a family of curves $\text{cos}(\delta/2)=\text{cos}(\Theta/2)\text{cos}(\Phi)$ with every $\delta/\pi\in\mathbb{Q}~(\bar{\mathbb{Q}})$ and plot regions of $A$ in Fig.~\ref{f3}(b). The diagram indicates the valid range of Eq. \ref{eq6}, which only converges in $\bar{A}$ when the system is ergodic. Thus, curves in Fig.~\ref{f2}(b) should be discontinuous everywhere in $\Phi$  (except for $\Theta=0, \pi$). They are a class of bizarre functions that are integrable (areas below the curve is well-defined), but derivatives diverge everywhere. 
 
For (iii), it is to break the illusion that $A$ and $\bar{A}$ are of ``equal" status, since (ii) has stated that they are both infinitely dense. In fact, $A$ and $\bar{A}$ correspond to different probability weight, namely \textit{measure}. When they are simultaneously encountered, only one subset is dominant in probability. If $D=1$, the measure of $\bar{A}$ is vanishing, i.e., probability of pumping is 0 (theorem 2.2(a)). If $D{\geq}2$, oppositely, the stable $m(A)=0$ and pumping surely occurs (theorem 2.2(b)).

The ``dephasing" associated with geometric pumping is a different type. First, the non-returning is not from ``real" dissipation, but from the journey becoming infinitely long. Second, non-returnable trajectories are statistically more than returnable ones, which has a geometric origin that non-close trajectories allowed by geometry will explode in number as dimension increases ($D>1$). Since the cardinality of $A$ proves equal to $\mathbb{Q}$ (Appx. C), one can also say dephasing is due to $\mathbb{Q}$ being sparser than $\bar{\mathbb{Q}}$. Thus, without fixing physical quantities precisely on $\mathbb{Q}$ or $\bar{\mathbb{Q}}$, two properties of $\mathbb{Q}$ and $\bar{\mathbb{Q}}$ still impose an influence: in \textit{arbitrary finite} interval (1) $m(\mathbb{Q})=0$; (2) both $\mathbb{Q}$ and $\bar{\mathbb{Q}}$ are dense subsets of $\mathbb{R}$.
 

It is easy to extend to a band model, in which geometric pumping still occurs. Consider two-band (spin-less) $H=\sum_i d_i(\textbf{k}; \textbf{R}(t))\sigma_i$, where $\sigma_i$ represents pseudo-spins (real spin can be restored by an expanded term $\sigma_i{\otimes}s_j$ \cite{BandM}); $d_i$ is controlled by parameter \textbf{R}(t) (e.g., time-dependent phonon amplitudes). Definition of topological invariants varies with symmetries and dimensions \cite{Symm, FieldTI, FuInv}, and here we just use an easy model to reveal the link between $\Theta$ and TPT. We adopt $d_x(k)=v+w{\cdot}\text{cos}(k{\cdot}l), d_y(k)=w{\cdot}\text{sin}(k{\cdot}l)$. This model can represent a chain (lattice $l$) with two atomic sites $A, B$ in each cell: $H=v\sum_{i}c^{\dagger}_{i, A}c_{i, B}+ w\sum_{i}c^{\dagger}_{i+1, A}c_{i, B}+h.c.$ Here, topological invariant is a $Z_2$ type, which is the winding number (1 or 0) of the circle spanned by $k$ around the origin of $(d_x, d_y)$ space \cite{Note2}. In resemblance to spin in Fig.~\ref{f1}(a), we can set $v(t)=a+\text{cos}({\omega}t), w(t)=1$. Evidently, $d_i$ plays the role of $B_i$ and the winding number will change at $\Theta=\pi$, and the trivial gap-close happens at $\Theta=0$. In general, a different choice of $v(t), w(t)$ (details of phonon's effects on bands) will change ``orientation" of $\textbf{B}$ fields, but not alter the angle $\Theta$.  Consequently, $\Theta$ in band models is quantized, taking values in two infinitesimal regions around 0 and $\pi$ \cite{Note3}. Basically, we have neglected inelastic phonon scattering and taken each $k$ as an independent spin model. Since gap close/opening is for a specific $k$, pumping may apply to node change in semi-metals. The time-reversal transformation $\mathcal{T}H\mathcal{T}^{-1}=-H$ used in derivation of Eq.~\ref{eq3} should be replaced by particle-hole symmetry ${\Xi}H{\Xi}^{-1}=-H$. The finite $m$ still satisfies in the new context. 
\begin{figure}
\includegraphics[scale=0.40]{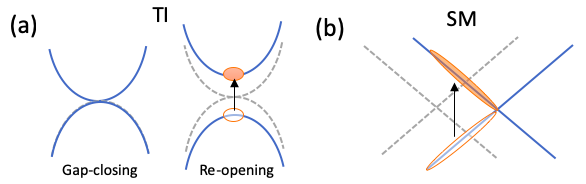}
\caption{\label{fig:epsart}(color online): Geometric pumping casued by band evolution in (a) topological insulator, (b) semi-metal. The pumping regions are highlighted. \label{f4}}
\end{figure}

We suggest observing geometric pumping in the vicinity of various TPT \cite{TPT} (e.g., narrow-gap topological insulator \cite{Chirag, Liang} or semi-metals \cite{Weyl}). Under the influence of phonons, the gap is closed and opened, causing periodic TPT; or a semi-metal's node position is changed (Fig.~\ref{f4}). Recently, a mystique of charge pumping has been detected in ZrTe$_5$ \cite{Chirag}, where carriers hop into upper bulk bands with below-gap pumping. The proposed geometric pumping provides a compelling implication to establish a general framework for light-topology quantum control experiments \cite{Luo, Yang, Photo, Liu, 1, 2, 3}.

This work reports a pumping caused by TPT, challenging the wisdom that $Z_2$ only causes surface observables. The pumping probability is exactly solved (Eqs.~\ref{eq6}, \ref{eq7}), showing a upper limit $\frac{1}{2}$, matching the numerical results (Fig.~\ref{f2}(b)). Its geometric feature (Eq.~\ref{eq7}) is salient. The conditions of pumping are proved (theorems 2.1 2.2). The measure-preserving formalism defined by Eqs.~\ref{eq4}, \ref{eq5} represents a route to incorporate geometry other than Berry's scheme. The rational/irrational numbers have entered physics when dimension $D>1$, and an intriguing question is if entropy can be defined linked to $\mathbb{Q}$ or $\bar{\mathbb{Q}}$.

\input acknowledgement.tex   

\section{Appendix}
\subsection{A. Derivation of evolution matrix $\mathcal{U}$.}
Although $\mathcal{U}$ is not adiabatic, evolution $U_c$ (the arc section) is. In the adiabatic limit, spin is rigorously aligned with \textbf{B} field, thus $U_{c}$ can be expressed in rotation angles of \textbf{B} field. $U_{c}=\mathcal{R}{\cdot}{\Lambda}$, where $\mathcal{R}(\alpha, \beta, \delta)$ are rotations of $\delta$ with an axis of polar and azimuthal angles $\alpha$, $\beta$. 
\begin{equation}
\begin{split}
\begin{pmatrix} 
\text{cos}(\frac{\delta}{2})-i\text{sin}(\frac{\delta}{2})\text{cos}(\alpha) & -i\text{sin}(\frac{\delta}{2})\text{sin}\alpha{\cdot}e^{-i\beta} \\[4pt]
-i\text{sin}(\frac{\delta}{2})\text{sin}\alpha{\cdot}e^{i\beta} &  \text{cos}(\frac{\delta}{2})+i\text{sin}(\frac{\delta}{2})\text{cos}(\alpha) \end{pmatrix}
\end{split}
\label{eq8}
\end{equation}
$\mathcal{R}$ will align spin with \textbf{B}. Based on the setting of Fig.~\ref{f1}(b), $\mathcal{R}$ has $\alpha=\pi/2, \beta=\Omega+\pi/2, \delta=\Theta$. ${\Lambda}$ is a diagonal matrix due to the undetermined phase. 
\begin{equation}
\Lambda=\begin{pmatrix} e^{-i\Gamma(\alpha, \beta, \delta)} & 0 \\ 0 & e^{i{\Gamma}'(\alpha, \beta, \delta)} \end{pmatrix},
\label{eq9}
\end{equation}
where $\Gamma$ and $\Gamma'$ are functions of rotation angles. In general, $\mathcal{R}{\cdot}{\Lambda}$ is a U(2) matrix. In this instance, a constraint narrows it down to SU(2). Time-reversal $\mathcal{T}=-i\sigma_yK$ and $K$ is complex conjugation. Then $\mathcal{T}U_{c}\mathcal{T}^{-1}$ is
\begin{equation}
\begin{split}
\mathcal{T}&[\mathfrak{T}~e^{-i{\int}^t_0H(\tau)d\tau}]\mathcal{T}^{-1}=\mathbb{I}+i\int_0^t\mathcal{T}H(\tau)\mathcal{T}^{-1}d\tau \\
&+i^2\int_0^t\int_0^{\tau}\mathcal{T}H(\tau)\mathcal{T}^{-1}\mathcal{T}H(\tau')\mathcal{T}^{-1}d{\tau}d{\tau}'+... \\
&=\mathfrak{T}~\text{exp}(i{\int}^t_0\mathcal{T}H(\tau)\mathcal{T}^{-1}d\tau).
\end{split}
\label{eq10}
\end{equation}
Here, $\mathfrak{T}$ denotes time-ordering; $H(t)=-{\sum}B_i(t)\sigma_i$, we have $\mathcal{T}H(t)\mathcal{T}^{-1}=-H(t)$.  Using this property and also $U_c=\mathcal{R}{\cdot}{\Lambda}$, we deduce
\begin{equation}
\begin{split}
\mathcal{T}U_{c}\mathcal{T}^{-1}=U_c,~\Gamma=\Gamma'.
\end{split}
\label{eq11}
\end{equation}
$\Gamma=\Gamma'$ will narrow down $\mathcal{U}$ to a SU(2). Then, the loop evolution operator $\mathcal{U}$ is the product of straight sections $U_1, U_2$ and the arc section $U_c$
\begin{equation}
\begin{split}
\mathcal{U}=U_{2}U_{c}U_{1}=\mathcal{R}\mathcal{R}^{-1}U_2\mathcal{R}\mathcal{R}^{-1}(\mathcal{R}{\Lambda})U_1.
\end{split}
\label{eq12}
\end{equation}
Note $\mathcal{R}^{-1}...\mathcal{R}$ will transform to bases $|n_{0,1}{\rangle}$ of $(\text{cos}(\frac{\Theta}{2}), \text{sin}(\frac{\Theta}{2})e^{i{\Omega}})^{T}$ and $(-\text{sin}(\frac{\Theta}{2})e^{-i{\Omega}}, \text{cos}(\frac{\Theta}{2}))^{T}$ (the inevitable stringent points are put on the S-pole for both branches). With the new bases, $U_{2}'=\mathcal{R}^{-1}U_2\mathcal{R}$ is diagonalized. ($U_1$ is already diagonal)
\begin{equation}
\begin{split}
U_2'=\begin{pmatrix} e^{-i\Phi_2} & 0 \\ 0 & e^{i\Phi_2} \end{pmatrix},~ U_1=\begin{pmatrix} e^{-i\Phi_1} & 0 \\ 0 & e^{i\Phi_1} \end{pmatrix},
\end{split}
\label{eq13}
\end{equation}
Define variables $\Phi_c=\Gamma$, $\Phi=\Phi_1+\Phi_2+\Phi_c$. Plug them in to Eq.~\ref{eq12}, we obtain
\begin{equation}
\begin{split}
\mathcal{U}=\mathcal{R}U_2'{\Lambda}U_1=\mathcal{R}(\frac{\pi}{2}, \Omega+\frac{\pi}{2}, \Theta)\begin{pmatrix} e^{-i\Phi} & 0 \\ 0 & e^{i\Phi} \end{pmatrix},
\end{split}
\label{eq14}
\end{equation}
which is just Eq.~\ref{eq3}.

\subsection{B. Ergodicity, probability space, Haar measure}
Here we give a physical view about definition 1. we can interpret $B$ with $m(B)>0$ as the initial setting of an ensemble. For each step of evolution $T$, we add the newly covered region to $B$, obtaining ${\bigcup}_{i=0}~T^{-i}B$. After sufficient many steps, an invariant $T^{-1}B=B$ indicates the maximum region of the phase space that may be reached. Therefore, $m(B)=1$ implies the system can eventually cover every part of the phase space, i.e., either right on or arbitrarily close to any point. An equivalent definition of ergodicity is, for instance (Theorem 1.5 of \cite{Ebook}):
\vskip 1.0mm
If $T: X{\to}X$ is a measure-preserving transformation of the probability space $(X, \mathfrak{B}, m)$, being ergodic is that for every $A$, $B{\in}\mathfrak{B}$ with $m(A)>0$, $m(B)>0$, there exists $n>0$ with $m(T^{-n}A~{\cap}~B)>0$.
\vskip 1.0mm
Since $A$, $B$ are arbitrary, starting from any region $A$, one can reach arbitrary region $B$ with sufficient steps $n$. Thus the choice of the initial state makes no difference to an ergodic system when distant past and future are included. ``Reach" means arbitarily close to. Thus it requires $m(A)>0$ and $m(B)>0$ for some error tolerance (allowing unreachable points, but not areas). If $A=B$, it is called ``reccurence", a weaker property enjoyed by all measure-preserving transformations, known as Poincar\' e recurrence theorem (Theorem 1.4 of \cite{Ebook}). Approximate or conditional measure preservation, like Wigner flow, will prevent applying theorems developed in ergodic theory. Thus, it is crucial to establish rigorous MPF on a suitable measure space.

The pair $(X, \mathfrak{B})$ is called a \textit{measurable space}. $\mathfrak{B}$ is a $\sigma$-algebra of subsets of $X$ satisfying (i) $X{\in}\mathfrak{B}$, (ii) if $B{\in}\mathfrak{B}$, $X/B{\in}\mathfrak{B}$, (iii) if $B_n{\in}\mathfrak{B}$ then ${\bigcup}_{n=1}^{\infty}B_n{\in}{\mathfrak{B}}$. \textit{Measure function} on $(X, \mathfrak{B})$ is defined as a mapping $m: \mathfrak{B}{\to}\mathbb{R}^{+}$ satisfying $m(\emptyset)=0$ and $m({\bigcup}_nB_n)={\sum}_{n}m(B_n)$, where $B_n$ is a sequence of members of $\mathfrak{B}$ which are pairwise disjoint subsets of $X$. A \textit{measure space} is a triple $(X, \mathfrak{B}, m)$, where $(X, \mathfrak{B})$ is a measurable space and $m$ is a finite measure function on it. $(X, \mathfrak{B}, m)$ is a \textit{probability space} if $m(X)=1$, and $m$ is called a \textit{probability measure}.
\vskip 1.0mm
\textit{Haar measure} is a probability measure on a compact group $G$ which ties with the group structure on $G$ (Sec. 0.6 of \cite{Ebook}). Simply speaking, it is a probability measure invariant under all group transformations and can be proved unique (Theorem 0.3 of \cite{Ebook}). Intuitively, one can imagine a measure function as a density field, and Haar measure is just a ``uniform" density. It can be conveniently represented in a differential form. For SU(2) group parameterized by Euler angles, $m(\phi, \theta, \psi)=\text{sin}(2\theta){\cdot}d{\phi}{\cdot}d{\theta}{\cdot}d{\psi}$. The extra factor ``2" in the main text arises from variable substitution. For 1D rotation with a fixed axis, Haar measure is simply $m(\eta)=d{\eta}$ for Fig.~\ref{f2}(a). Formally, Haar measure is defined as $m(E')=m(E)$ for $E'=xE$ and $E'=Ex$, ${\forall}x{\in}G$, ${\forall}E{\in}\mathfrak{B}$. Rigorously, $E$ belongs to the $\sigma$-algebra $\mathfrak{B}$, but from loose point view, $E$ is just a subset of group $G$, and $E'=xE$ or $Ex$ is just rotating $E$ by an operation $x$ in $G$. Then $m$ remains invariant under all these rotations. 

In this case, $x$ corresponds to unitary evolution operator $\mathcal{U} {\in}$ SU(2). By the defining features of Haar measure, we have $m(E)=m(\mathcal{U}E)=...=m(\mathcal{U}^{n}E)$, where $E$ can either stand for a single initial state or a collection of initial states (when dealing with an ensemble). $m(E)=m(\mathcal{U}^{n}E)$ is to replace its classical counterpart $m(E)=m(T_s^n(p, q)E)$. The current MPF setting is not limited to SU(2), making for additional options of other continuous groups, as it merely rests on a math fact that continuous surjective endomorphism of a compact group will preserve its Haar measure. We will not give proof here but emphasize, in terms of physics application, most groups of interest are compact, such as SU$(n)$, SO$(n)$, U$(n)$, etc. The evolution operator is bijective (physically, bijectivity corresponds to the fact that the state's future and past must be uniquely determined by Schr\"odinger equation), thus surjectivity is satisfied too. 


\subsection{C. Ergodic subgroup of SU(2)}
An ergodic group must be abelian (Theorem 1.1). First, we find an abelian subgroup of SU(2), which is just the group formed by rotations with a fixed axis $\mathcal{R}(\alpha, \beta, \delta)$ with $\alpha$, $\beta$ specifying the axis orientation and $\delta$ being the rotation angle. It satisfies $\mathcal{R}(\alpha, \beta, \delta)\mathcal{R}(\alpha, \beta, {\delta}')=\mathcal{R}(\alpha, \beta, {\delta}+{\delta}')$. For example, if we rotate with $x$-axis by $\pi/3$ counter-clockwise, we have $\mathcal{R}(\frac{\pi}{2}, 0, \frac{\pi}{3})$. 

In addition, to be ergodic, ${\lbrace}\mathcal{R}^n|n{\in}\mathbb{N}{\rbrace}$ should be dense (Theorem 1.1), which is true iff $2{\pi}/{\delta}{\in}\bar{\mathbb{Q}}$ or ${\delta}/{\pi}{\in}\bar{\mathbb{Q}}$. Thus, the ergodic subgroup $G$ is 
\begin{equation}
G={\lbrace}\mathcal{R}(\alpha, \beta, \delta)|~{\delta}/{\pi}{\in}\bar{\mathbb{Q}}{\rbrace}.
\label{eq15}
\end{equation}

Comparing with SU(2) matrices parameterized by Euler angles $(\phi, \theta, \psi)$
\begin{equation}
R(\phi, \theta, \psi)=\begin{pmatrix} \text{cos}(\frac{\theta}{2})~e^{-i\frac{\phi+\psi}{2}} & -i{\cdot}\text{sin}(\frac{\theta}{2})~e^{i\frac{\psi-\phi}{2}} \\ -i{\cdot}\text{sin}(\frac{\theta}{2})~e^{i\frac{\phi-\psi}{2}} & \text{cos}(\frac{\theta}{2})~e^{i\frac{\phi+\psi}{2}}   \end{pmatrix},
\label{eq16}
\end{equation}
we obtain the mapping $(\alpha, \beta, \delta){\leftrightarrow}(\phi, \theta, \psi)$:
\begin{equation}
\begin{split}
&\text{cos}({\delta}/2)=\text{cos}(\theta/2)\text{cos}(\frac{\phi+\psi}{2}), \\
&\text{cos}^2{\alpha}=\frac{\text{cos}^2(\theta/2)\text{sin}^2(\frac{\phi+\psi}{2})}{1-\text{cos}^2(\theta/2)\text{cos}^2(\frac{\phi+\psi}{2})}, \\
&\beta=\frac{(\phi-\psi)}{2}+n\pi, n=0, 1.
\end{split}
\label{eq17}
\end{equation}
Comparing $R(\phi, \theta, \psi)$ with $\mathcal{U}$, we obtain the mapping of $(\phi, \theta, \psi){\leftrightarrow}(\Theta, \Phi, \Omega)$:
\begin{equation}
\begin{split}
\phi=2{\Phi}+{\Omega}-\frac{\pi}{2},~\theta={\Theta},~\psi=-\Omega+\frac{\pi}{2}.
\end{split}
\label{eq18}
\end{equation}
Then, we can deduce two results. The first is Eq.~\ref{eq6}. In Fig.~\ref{f2}(a), spin is located by angles $\eta$ and $\alpha$. Denote the angle between spin and $z$-axis is $\zeta(\eta, \alpha)$, which can be found by a geometry relation $\text{sin}(\zeta/2)=\text{sin}(\alpha){\cdot}\text{sin}(\eta/2)$. The projection (pump) probability at a given orientation is $\text{sin}^2(\zeta/2)$. $p_{\infty}$ is the following integral over the trajectory (Fig.~\ref{f2}(a)), and probability density $\rho$ should be constant by Theorem 1.2.
\begin{equation}
\begin{split}
p_{\infty}=\frac{{\int}_0^{2\pi}\text{sin}^2(\zeta/2){\cdot}{\rho}~d{\eta}}{\int_0^{2\pi}{\rho}~d{\eta}}=\frac{1}{2}\text{sin}^2(\alpha).
\end{split}
\label{eq19}
\end{equation}
Plug Eq.~\ref{eq17}, \ref{eq18} into Eq.~\ref{eq19}, we obtain Eq.~\ref{eq6}.

The second result is the distribution of $A$ in phase space (Fig.~\ref{f3}(b)). With Eq.~\ref{eq17}, \ref{eq18}, we can express $\delta$ in terms of $\Phi$, $\Theta$. Since $\delta{\in}\mathbb{Q}$, the stable set $A$ is
\begin{equation}
\begin{split}
{\lbrace}(\Theta, \Phi)|~\text{cos}(\Theta/2){\cdot}\text{cos}({\Phi})=\text{cos}({\delta/2}), {\delta}/{\pi}{\in}\mathbb{Q}{\rbrace}
\end{split}
\label{eq20}
\end{equation}
From Above, we immediately realize that the $A$ has the same cardinality as $\mathbb{Q}$ (thus, $\bar{A}$ has the same cardinality as $\bar{\mathbb{Q}}$). That means the lines in Fig.~\ref{f3}(b) should be as many as rational number, a set of supper dense curves.


\subsection{D. Proof of theorems 2.1, 2.2}
\textit{Proof}. By the Cayley–Hamilton theorem, the matrix $\mathcal{U}$ is a root of its characteristic polynomial $X^2-X\mathrm{tr}\,\mathcal U+\det\mathcal U$ in matrix space. Thus we have
\begin{equation}
\begin{split}
\mathcal U^2=2\cos(\Theta/2)\cos(\Phi)\cdot\mathcal U-\mathbb{I}.
\end{split}
\label{eq21}
\end{equation}
Substituting $\mathcal{U}=i~\tilde{\mathcal{U}}$ yields
\begin{equation}
\begin{split}
\tilde{\mathcal U}^{2}=\tilde\lambda~\tilde{\mathcal U}+\mathbb{I}, ~~\tilde{\mathcal U}^{n+2}=\tilde{\lambda}~\tilde{\mathcal U}^{n+1}+\tilde{\mathcal U}^n
\end{split} 
\label{eq22}
\end{equation}
with
\begin{equation}
\tilde\lambda
=-i\,\mathrm{Tr}(\mathcal U)
=-2i\cos(\Theta/2)\cos(\Phi)\,,
\label{eq23}
\end{equation}

Taking an indeterminate $X$, the \emph{Fibonacci polynomials} $F_n(X)$ and $E_n(X)$ are defined recursively by
\begin{equation}
P_{n+2}(X)=XP_{n+1}(X)+P_n(X)
\label{eq24}
\end{equation}
with respective recursion bases
$$
F_0=0\,,\ F_1=1
\
\mbox{ and }
\
E_0=1\,,\ E_1=0\,.
$$
Thus $E_{n}=F_{n-1}$ for $n\in\mathbb N$. For example, $F_2(\tilde{\lambda})=\tilde{\lambda}$, $F_3(\tilde{\lambda})=\tilde{\lambda}^2+1$, $F_4(\tilde{\lambda})=\tilde{\lambda}^3+2\tilde{\lambda}$, etc. Induction with Eq.~\ref{eq24} confirms the expression
\begin{equation}
\tilde{\mathcal U}^{n}=F_n(\tilde\lambda)\cdot\tilde{\mathcal U}+E_n(\tilde\lambda)\cdot\mathbb I
\label{eq25}
\end{equation}
Being stable ${\Leftrightarrow}~(\mathcal{U}^{n})_{1,2}=0$. The off-diagonal term is
\begin{equation}
(\mathcal{U}^{n})_{1,2}=i^{n+1}F_n(\tilde{\lambda})~e^{i{\Phi}}\text{sin}(\Theta/2).
\label{eq26}
\end{equation}

Let $\mathcal F_n$ be the finite set of roots of all the Fibonacci polynomials of degree up to $n$. Then the countable set $\mathcal F=\lim_{n\to\infty}\mathcal F_n$ is dense in $2i[-1, 1]$. More formally stated: as $n$ tends to infinity, the set $\mathcal F_n$ converges to $2i[-1, 1]$ in the Hausdorff metric on sets [Th.~1.1] of J. Number Theory, \textbf{163}, 89-100 (2016).
\vskip 1mm
\noindent
Theorem 2.1(a): If $\Theta=0$ or $\pi$, $\text{sin}(\Theta/2)=0$ and $(\tilde{\mathcal{U}}^{n})_{1,2}=0$. Thus these points are stable, independent of $\Phi$. If $\Phi={\pm}\pi/2$, we have $\tilde\lambda=0$. We also have $F_{2n}(0)=0$, they are all stable points of order 2 independent of $\Theta$.
\vskip 1mm
\noindent
Theorem 2.1(b): 
Fix $\Theta\in(0,\pi)$. In particular, note $\cos\frac\Theta2>0$. Now consider $\varphi\in(-\pi/2,\pi/2)$ and $\varepsilon>0$. Since the inverse cosine function is continuous at $\cos\varphi$, there is a number $\delta>0$ such that $|\cos\varphi-\cos\Phi|<\delta$ implies $|\varphi-\Phi|<\varepsilon$. Since $2i\cos\frac\Theta2\cos\varphi\in2i[-1,1]$ and the set $\mathcal F$ of roots of Fibonacci polynomials is dense in $2i[-1,1]$, there is a root $2i\cos\frac\Theta2\cos\Phi\in2i[-1,1]$ with
$$
|\cos\tfrac\Theta2\cos\varphi-\cos\tfrac\Theta2\cos\Phi|<\delta\cos\tfrac\Theta2\,.
$$
Thus $(\Theta, \Phi)\in A$ for $|\varphi-\Phi|<\varepsilon$, as required for (i). Finally, since $\Set{\Phi|(\Theta, \Phi)\in A}$ is a countable subset of $[-\pi/2,\pi/2]$, its complement $\Set{\Phi|(\Theta, \Phi)\in\bar A}$ is a dense subset, yielding (ii).
\vskip 1mm
\noindent
Theorem 2.1(c): is similar to Theorem 2.1(b).
\vskip 1mm
\noindent
Theorem 2.2(a): follows from Theorem 2.1(a).
\vskip 1mm
\noindent
Theorem 2.2(b): Since stable points correspond to roots of Fibonacci polynomials, there are only countably many of them, and they form a set of measure zero in the 2D or 3D phase spaces. Note that stability is independent of $\Omega$, which trivially increases the dimension by one. $\square$

\end{document}

%% file: acknowledgement.tex
\textbf{Acknowledgement}. This work was supported by the Ames Laboratory, the US Department of Energy, Office of Science,
Basic Energy Sciences, Materials Science and Engineering Division under contract No. DEAC02-07CH11358. Terahertz instrument was supported in part by the National Quantum Information Science Research Center; Superconducting Quantum Materials Systems Center.

%% file: Ad_Pump_8.bbl
\begin{thebibliography}{99}
\bibitem{MPF} Measure preservation means a conservative system's volume in phase space ${\lbrace}p, q{\rbrace}$ is constant during evolution, known as the Liouville theorem (Ref. \cite{Statistics}). However, its quantum counterpart, Wigner flow (Ref. \cite{Wigner, Flow}), is compressible, which depreciates usage and poses a long-pending question: if measure preserving formalism can be rigorously constructed and significantly used in quantum scenarios (Ref. \cite{Wigner, Gro46, Moy49}).
\bibitem{Note} Note that Eq.~\ref{eq7} only converges in $\bar{A}$, but ${\lbrace}({\Theta}, {\Phi})|\Theta=\pi{\rbrace}{\subset}A$ (Theorem 2.1(a)). It is important to remember $\Theta=\pi$ is a short notation for ``$\Theta=\pi$ and its infinitesimal neighborhood". Physically, it means presence of fluctuation, which is always satisfied in real materials.
\end{thebibliography}
